\documentclass[letterpaper,twoside,twocolumn,english]{revtex4}
\usepackage[T1]{fontenc}
\usepackage[utf8]{inputenc}
\usepackage{amsmath}
\usepackage{amssymb}
\usepackage{graphicx}
\usepackage{esint}
\usepackage{babel}

\begin{document}

\title{Effects of spontaneous symmetry break in the origin of non-analytic behavior of
entanglement at quantum phase transitions}

\author{L. Justino and Thiago R. de Oliveira}

\affiliation{Instituto de Física, Universidade Federal Fluminense, Av. Gal. Milton
Tavares de Souza s/n, Gragoatá, 24210-346, Niterói, RJ, Brazil}

\begin{abstract}
We present an example where Spontaneous Symmetry Breaking may effect not only the behavior of the entanglement at Quantum Phase Transitions, but also the origin of the non-analyticity. 
In particular, in the XXZ model, we study the non analyticities in the concurrence between two spins, which was claimed to be accidental, since it had its origin in the optimization involved
in the concurrence definition. We show that when one takes in account the effect of the Spontaneous
Symmetry Breaking, even tough the values of the entanglement measure does not change, the origin the the non-analytical behavior changes: it is not due to the optimization process anymore and in this sense
it is a "natural" non-analyticity. This is a much more subtle influence of the Spontaneous Symmetry Breaking not noticed before. However the non-analytical behavior still suggests a second order quantum phase transition and not the first order that occurs and we explain why. We also show that the value of entanglement between one site and the rest of the chain does change when taking into account the Spontaneous Symmetry Breaking.
\end{abstract}

\maketitle

\section{{\normalsize Introduction}}

It is now generally accepted that entanglement may help in finding and characterizing Quantum Phase
Transitions (QPT), since it may inherit the non-analytic behavior of the ground state energy 
\cite{Wu04, Wu06} (see \cite{Amico} for a review on entanglement and QPT). However such a relation has its caveats. There are non-analytical behavior in 
entanglement measurements which do not correspond to QPT, as first showed by Min-Fong Yang \cite{Yang05}, 
for example. In general, that happens because entanglement measures are defined using optimization 
procedures which may create accidental non-analytical behavior or even hidden genuine ones.

Besides such caveats, the study of entanglement at QPT is in general more intricate than of 
thermodynamical quantities, because of the Spontaneous Symmetry Breaking (SSB). At the critical point 
of a symmetry breaking QPT, the ground state becomes degenerate. In fact, this degenerescence is 
necessary for the spontaneous breaking of the Hamiltonian symmetry: the emergence of ground 
states without the Hamiltonian symmetry. However, as the ground state is degenerate, there 
are many possible ground states; some preserving the symmetry (equal superpositions of
symmetry breaking states) others not. And, while thermodynamical 
quantities do not depend on which particular degenerate ground state one choose, entanglement 
do. Therefore one has to be careful when choosing the state. In general, states which preserve 
the symmetry are preferable, since they are simpler: have reduced density matrices with many 
null entries. And there are examples where the entanglement does not change \cite{Syljuasen}, but 
also where it does \cite{Oliveira,Osterloh06}.

Here, we show an example of a new and very subtle caveat in the relation between entanglement, SSB and QPT.
More  specifically, we show that the SSB may change not only the value of the entanglement but also
the origin of its non-analytical behavior. Actually, in our example, SSB changes the origin of the
non-analytical behavior wihtout even changing the value of the entanglement; a much more subtle influence.

\section{{\normalsize Spin-$\frac{1}{2}$ XXZ model}}

We will first introduce the model we study. The one dimensional spin-$\frac{1}{2}$ XXZ chain is given by the following Hamiltonian
\begin{equation}
H=\sum_{i=1}^{\infty} (\sigma_i^x \sigma_{i+1}^x + \sigma_i^y \sigma_{i+1}^y + \Delta \sigma_i^z \sigma_{i+1}^z),
\end{equation}
where $S_{j}^{u}=\sigma_{j}^{u}/2$ ($u=x,y,z$), $\sigma_{j}^{u}$
are the Pauli spin-$\frac{1}{2}$ operators on site $j$, $\Delta$
is the anisotropy parameter, and we consider periodic boundary
conditions: $\sigma_{j+N}^{u}=\sigma_{j}^{u}$.
The XXZ model cannot be diagonalized, but its energy spectrum can
be obtained by Bethe ansatz. The Hamiltonian has two symmetries: i)
a discrete parity $\mathbb{Z}_{2}$ symmetry over the plane $xy$:
$\sigma^{z}\rightarrow-\sigma^{z}$, and ii) a continuous $U(1)$
symmetry that rotates the spins in the $xy$ plane by any angle $\theta$
. The $\mathbb{Z}_{2}$ symmetry implies that $\langle\sigma_{i}^{z}\rangle=0$
and $\langle\sigma_{i}^{x}\sigma_{j}^{z}\rangle=\langle\sigma_{i}^{y}\sigma_{j}^{z}\rangle=0$,
while the $U(1)$ symmetry implies that $\langle\sigma_{i}^{x}\rangle=\langle\sigma_{i}^{y}\rangle=0$,
$\langle\sigma_{i}^{y}\sigma_{j}^{y}\rangle=\langle\sigma_{i}^{x}\sigma_{j}^{x}\rangle$
and $\langle\sigma_{i}^{x}\sigma_{j}^{y}\rangle=0$. The model
has three phases, but we will analyze only two of then: 
\begin{description}
\item [{i)}] $\Delta\leq-1$: the system is in a ferromagnetic phase with
all the spins pointing in the same direction, which breaks the discrete $\mathbb{Z}_{2}$ 
symmetry, creating a finite magnetization $(\langle \sigma_j^z \rangle = \langle \sigma_i^z \rangle = m)$. 
The critical point at $\Delta=-1$ is of first order.
\item [{ii)}] $-1<\Delta<1$: the system is in a gapless phase, where the
correlation decays polynomially and all the symmetries are preserved.
\end{description}

The Bethe ansatz solution gives the ground state energy \cite{Shiroishi,Yang}:
{\small 
\begin{equation}
e_{0}(\Delta)=\begin{cases}
-\frac{\Delta}{4}, & \Delta\leq-1,\\
\frac{\Delta}{4}+\frac{\sin{\pi\nu}}{2\pi}\int_{-\infty+\frac{i}{2}}^{\infty+\frac{i}{2}}dx\frac{1}{\sinh{x}}\frac{\cosh{\nu x}}{\sinh{\nu x}} & ,-1<\Delta<1,
\end{cases}
\end{equation}
}where $\Delta=\cos{\pi\nu}$. For nearest neighbors we can obtain the correlation
from $e_{0}(\Delta)$:
\begin{equation}
\langle\sigma_{i}^{z}\sigma_{i+1}^{z}\rangle=4\frac{\partial e_{0}(\Delta)}{\partial\Delta},
\end{equation}
\begin{equation}
\langle\sigma_{i}^{x}\sigma_{i+1}^{x}\rangle=\langle\sigma_{i}^{y}\sigma_{i+1}^{y}\rangle=\frac{1}{2}(4e_{0}(\Delta)-\Delta\langle\sigma_{i}^{z}\sigma_{i+1}^{z}\rangle).
\end{equation}
For spins further apart, progress has been slow, but there are already
some expressions available up to third neighbors \cite{Shiroishi}.
We will not show them here, since they are too lengthy %
\footnote{Note that there are typos in equations (19) and (20) from \cite{Shiroishi}.
In (19) we only need to sum a $-\frac{c_{1}}{\pi s_{1}}\zeta_{\nu}$.
In (20) we need to go to \cite{Kato} (note that equation (5.4) has
the same typo) and use equations (5.10), (B.11) and (B.12) to calculate
and find the error in $\langle\sigma_{i}^{x}\sigma_{i+3}^{x}\rangle$%
}.

We are interested in the subtleties of SSB in entanglement measurements 
for two spins in this chain. These entanglement measurements can be determined by 
the reduced density matrix of the two spins, which can be obtained from the 
magnetizations and correlations of the two spins. Actually, the general density
matrix of two spins-$\frac{1}{2}$ can be expressed as {\small 
\begin{equation}
\rho_{i(i+r)}=\frac{1}{4}\left[\mathbb{I\otimes\mathbb{I}}+\mathbf{p.\sigma}\otimes\mathbb{I}+\mathbb{I}\otimes\mathbf{q.\sigma}+\sum_{u,v}t_{r}^{uv}\sigma_{i}^{u}\otimes\sigma_{i+r}^{v}\right],
\end{equation}
with} $r$ being the distance between the sites, $\sigma=(\sigma^x,\sigma^y,\sigma^z)$ , $\mathbf{p=\langle\sigma}\otimes\mathbb{I}\rangle$,
${\bf q}=\langle\mathbb{I}\otimes\mathbf{\sigma}\rangle$, and $t_{r}^{uv}=\langle\sigma_{i}^{u}\sigma_{i+r}^{v}\rangle$.
In the case of the XXZ model, due to the Hamiltonian symmetries, only $\{t_{r}^{xx},t_{r}^{yy},t_{r}^{zz}\}$ do not vanish and $t_{r}^{yy}=t_{r}^{xx}$. Translation invariance
also implies that the correlation functions for two-sites depend only
on the distance between the sites ($r$), being independent of $i$.
So, for the XXZ model we have 
\begin{equation}
\rho_{r}=\frac{1}{4}\left(\begin{array}{cccc}
1+t_{r}^{zz}  & 0 & 0 & 0\\
0 & 1-t_{r}^{zz} & 2t_{r}^{xx} & 0\\
0 & 2t_{r}^{xx} & 1-t_{r}^{zz} & 0\\
0 & 0 & 0 & 1+t_{r}^{zz}\\
\end{array}\right).
\end{equation}

\section{{\normalsize Concurrence and QPT}}

The first formal and general relation between entanglement and QPT was given in \cite{Wu04}. It is proved that: a discontinuity in or a divergence of the ground-state concurrence 
(the first derivative of the ground state concurrence) can be both necessary and sufficient condition to signal first-order QPT (second-order QPT), except in cases where the non-analyticity is artificial and/or accidental. In sum, the non-analyticity has to come from the matrix elements of the density matrix, not from the mathematical expression for the entanglement measure. At the same time, an explicit example of such an artificial non-analyticity was given for the concurrence of two spins in a XXZ chain \cite{Yang05}.

For the symmetric ground state the entanglement between two spins, given by the concurrence, can be obtained from the reduced density matrix and it has the simple formula:
\begin{equation} \label{A}
\mathcal{C}_r=\text{max}\{0,\tilde{\mathcal{C}}_r\}
\end{equation}
with 
\begin{equation} \label{C}
\tilde{\mathcal{C}}_r=\frac{1}{2}\left(2|\langle\sigma_x^i\sigma_x^{i+r}\rangle|-
(1+\langle\sigma_z^i\sigma_z^{i+r}\rangle\right).
\end{equation}
However if one consider the SSB, the expression for the concurrence is more complicated and 
given, as seen in \cite{Syljuasen}, by:
{\small
\begin{equation} \label{CSSB}
\tilde{\mathcal{C}}_r^{SSB}=\frac{1}{2}\left(2|\langle\sigma_x^i\sigma_x^{i+r}\rangle|-\sqrt{(1+\langle\sigma_z^i\sigma_z^{i+r}\rangle)^2-(p_z+q_z)^2}\right).
\end{equation}}
Note that the maximization in Eq. (\ref{A}) appears because the entanglement measure
involves an optimization procedure over all possible decomposition of the mixed
state in a mixture over pure states. There is also a modulus which comes from the
entanglement definition and could cause some accidental non-analyticity behavior (see \cite{Wootters} for more details on the definition
of the concurrence). Note also that, as expected, Eq. (\ref{CSSB}) reduce to Eq.~(\ref{C}) when there is no SSB ($p_z=q_z=m=0$).

In Fig.~(\ref{figconc}) we show $\tilde{\mathcal{C}}_r$, which is the concurrence without
taking into account the maximum operation neither the SSB. We can see that $\tilde{\mathcal{C}}_r$ is discontinuous at $\Delta=-1$ jumping from -1 to 0. This discontinuity has its origins in
$\langle\sigma_x^i\sigma_x^{i+r}\rangle$ and $\langle\sigma_z^i\sigma_z^{i+r}\rangle$, which are both discontinuous at $\Delta=-1$. A discontinuity in the concurrence would indicate a first-order QPT (1QPT).
But the true entanglement measure is $\mathcal{C}_r$ and not $\tilde{\mathcal{C}}_r$. The curves of 
$\mathcal{C}_r$ can be seen in Fig. (\ref{figconc}) if one just considers the positive 
vertical axis or in  {Fig.~(\ref{figconcsSSB}), since in this model the SSB does not 
change the value of the entanglement:  $\mathcal{C}_r=\mathcal{C}_r^{SSB}=\tilde{\mathcal{C}}_r^{SSB}$ 
(the last equality comes from the fact that $\tilde{\mathcal{C}}_r^{SSB} \geq 0$). One can see that
$\mathcal{C}_r$ is continuous, but changes abruptly. It is possible to check that the
first derivative of $\mathcal{C}_r$ is discontinuous (we checked it, but one can guess from
the form of the curve of $\mathcal{C}_r$), which should indicate an second-order
QPT (2QPT). In sum, the concurrence indicates an 2QPT, while it is known that at $\Delta=-1$
we have a 1QPT indicated by $\tilde{\mathcal{C}}_r$, which is not the 
true entanglement measure.

This failure of concurrence to indicate the right order of the QPT was noted in \cite{Yang05}
and happens because the discontinuity in the first derivative of $\mathcal{C}_r$ comes
from the maximum operation and not from the non-analytical behavior of the energy, which is 
present in the correlation functions. Then, this is an artificial non-analyticity and should
not indicate a QPT properly, as argued in \cite{Yang05}. However these results are obtained not taking 
in account the effects of the SSB: using Eq.~(\ref{C}) instead of Eq.~(\ref{CSSB}).

\begin{figure}[h]
\centering
\includegraphics[scale=1.0,keepaspectratio=true]{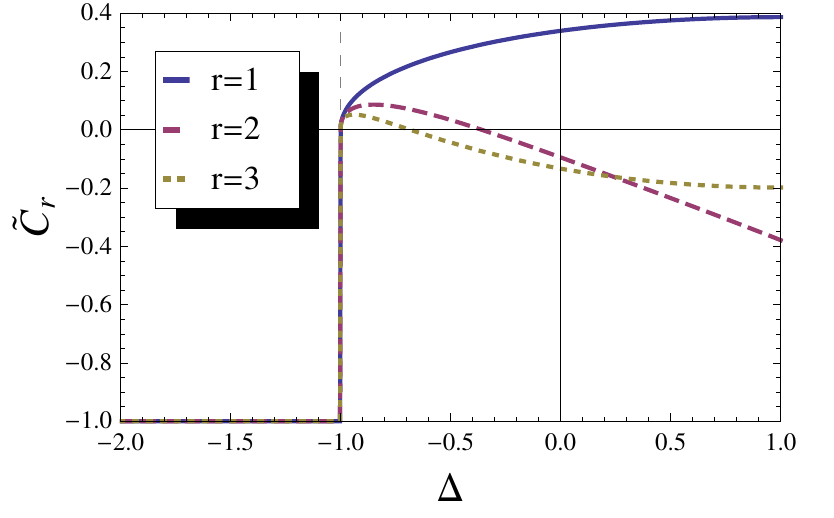}
\caption{\small{"Concurrence" before maximum operation for first, second and third neighbor
without taking into account the spontaneous symmetry breaking. We can see that the non analyticity
in the concurrence at $\Delta=-1$ 
is accidental, due to the maximum operation}}
\label{figconc}
\end{figure}

Lets consider now the effect of the SSB by taking into account that in the ferromagnetic phase, $\Delta<-1$, all spins are aligned in the same direction: $m=\pm 1$ ($p_z=q_z=m$). The expression
for the concurrence without taking into account the maximum operation is given by 
Eq.~(\ref{CSSB}), which is shown in Fig.~(\ref{figconcsSSB}). We can see that in the 
ferromagnetic phase $\tilde{\mathcal{C}}_r^{SSB}$ vanishes, so the concurrence goes 
to zero "naturally" without the maximum operation:
$\tilde{\mathcal{C}}_r^{SSB}=\mathcal{C}_r^{SSB}$. We have two facts here: 
1) even though the expressions are different
the entanglement value is the same taking or not in account the SSB, something already
noted by \cite{Syljuasen} (and by \cite{Osterloh06, Oliveira} for the XY model). 2) the origin of non-analyticity in $\mathcal{C}_r^{SSB}$
at $\Delta =-1$ is not due to the maximum operation, as it is in $\mathcal{C}_r$,
but comes from the correlation functions and the magnetizations. Therefore
we see that besides not changing the behavior of the concurrence the SSB does change
the origin of the non-analyticity turning it not an accidental but natural one.

Unfortunately $\mathcal{C}_r^{SSB}$ still indicate a 2QPT instead of a 1QPT.
That happens because the non-analytic behavior of the energy, which would indicate the 
correct 1QPT, is contained in the correlation function $\langle\sigma_z^i\sigma_z^{i+r}\rangle$, but
this is canceled by the term $p_z+q_z$ in Eq. (\ref{CSSB}). Thus in some sense
one could still argue that this is an accidental non-analytical behavior, but of 
different nature.

\begin{figure}[h]
\centering
\includegraphics[scale=0.85,keepaspectratio=true]{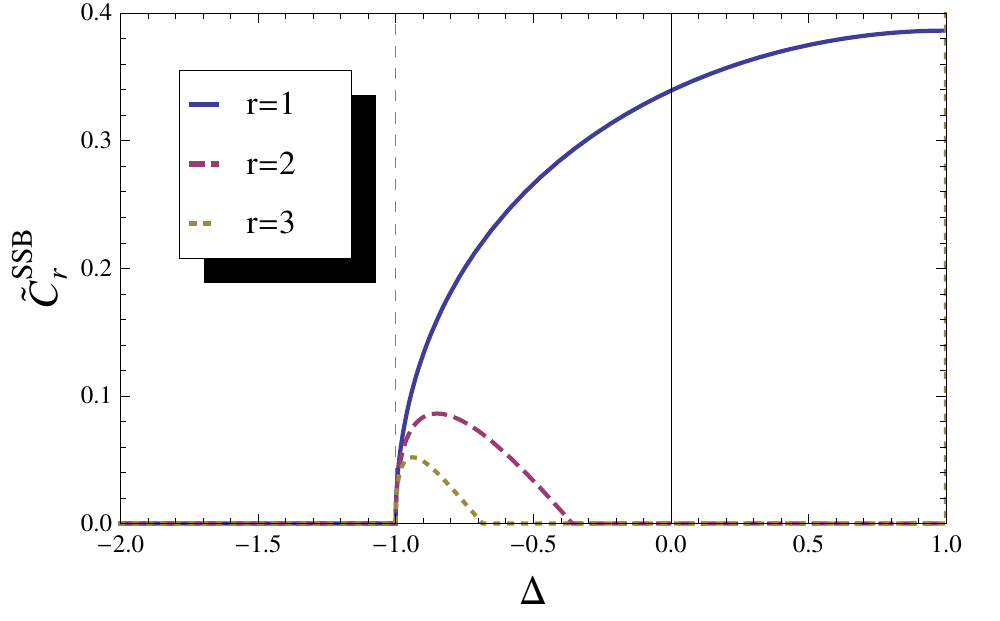}
\caption{\small{"Concurrence"\ before maximum operation for first, second and third neighbor 
taking into account the spontaneous symmetry breaking. We can see that the non analyticity in the concurrence at $\Delta=-1$, which was accidental due to the maximum operation, happens naturally if
we do take in account the symmetry breaking. Note that here "concurrence"\ is equal to concurrence in this case.}}
\label{figconcsSSB}
\end{figure}

\section{{\normalsize Von Neumann entropy and QPT}}

Another interesting fact is the raise of a discontinuity in the entanglement between one site and
the rest of the chain given by the von Neumann entropy for one site, when we take account the SSB.

The von Neumann entropy for one site is given by the equation:
\begin{equation} 
S=-x\log{x}-(1-x)\log{(1-x)},
\end{equation}
where $x$ is one of the two eigenvalues of the reduced density matrix of one site, that is
\begin{equation} 
\rho_i=\frac{1}{2}\left[\begin{array}{cc}
1+m_i  &  0\\
0  &  1-m_i
\end{array}\right],
\end{equation}
with $m_i=m=\langle\sigma_z\rangle$. So, our von Neumann entropy is:
{\small 
\begin{equation} 
S_i=-\left(\frac{1+m}{2}\right)\log{\left(\frac{1+m}{2}\right)}-\left(\frac{1-m}{2}\right)\log{\left(\frac{1-m}{2}\right)}.
\end{equation}}

\begin{figure}[h]
\centering
\includegraphics[scale=0.85,keepaspectratio=true]{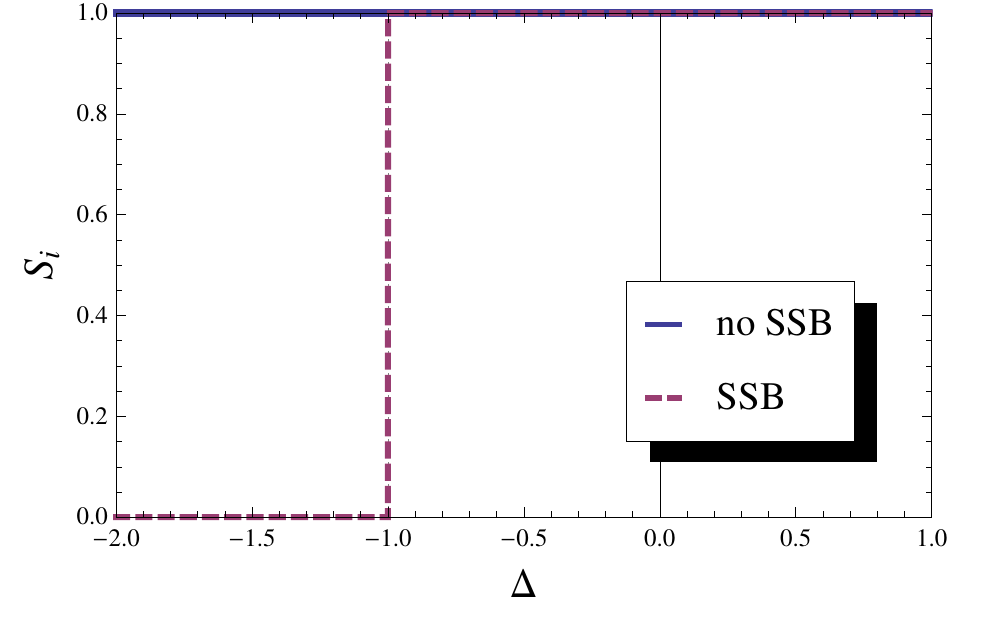}\caption{\small{Von Neumann entropy of particle $i$ with and without SSB, which shows the entanglement between 
this particle with all other particles in the chain. We can see that it can indicate 1QFT at $\Delta=-1$ if we take account SSB.}}
\label{figS}
\end{figure}

Fig.~(\ref{figS}) depicts this von Neumann entropy taking account the SSB, red dashed 
line, and without this SSB, blue line. It shows that without SSB this entropy indicates 
that the ferromagnetic phase is maximally entangled: it is a macroscopic superposition 
of two states with finite and opposite magnetization. Such state is very sensitive to 
external perturbations. Taking into account the SSB, the entanglement goes to zero as it 
should be for the separable ferromagnetic state at this phase; in this case the system 
chooses one of the two ferromagnetic configurations. Therefore, in this case the SSB 
influences directly in the entanglement behavior at the QPT, and has to be taken in account 
for the entanglement to signals the 1QPT. Such influence of the SSB has also been
found out in other models \cite{Osterloh06, Oliveira}.

\section{{\normalsize Conclusion}}

We have studied the influence of Spontaneous Symmetry Breaking in entanglement between two spins 
and between one spins and the rest of the chain in the one dimensional spin-$\frac{1}{2}$ XXZ 
model. We first showed that, although Spontaneous Symmetry Breaking does not change the behavior 
of the concurrence at the first-order Quantum Phase Transition, as first noted by \cite{Syljuasen}, 
it does changes the origin of non-analyticity behavior; a much more subtle influence. Yet, 
unfortunately, it still indicates a second-order Quantum Phase Transition instead of a first-order 
one.

We also showed that the behavior of the entanglement between one site and the rest is affected 
by the Spontaneous Symmetry Breaking at the first-order Quantum Phase Transition. It only signal 
the phase transition when taking into account the Spontaneous Symmetry Breaking.

Thus we give further evidence that the use of entanglement to study quantum phase transitions
may be much more intricate than at first glance. One has to be cautious about accidental
non-analytic behavior due to optimizations in the entanglement measure adopted, 
about the effects of the spontaneous symmetry breaking in the entanglement value, and
also to carefully take in account the effects of Spontaneous Symmetry Breaking in 
the origin of the non analiticities.

\begin{acknowledgments}
Both authors acknowledge financial support from the Brazilian agencies 
CNPq and CAPES. This work was performed as part of the Brazilian 
National Institute of Science and Technology for Quantum Information.
\end{acknowledgments}

\end{document}